\documentclass[aps,prl,paper,reprint,amsmath,amssymb,superscriptaddress,floatfix,showpacs,showkeys,subeqn,footinbib]{revtex4-1} 
\usepackage{amsmath, amsfonts, graphics, graphicx, hyperref,url,array, multirow, subfigure, rotating}
\usepackage{todonotes}
\usepackage{color}

\newcommand{\beq}{\begin{equation}}
\newcommand{\eeq}{\end{equation}}

\begin{document}

\title{The disc random packing problem: \\
a disorder criterion and an explicit solution}
\author{Raphael Blumenfeld}
\affiliation{Gonville \& Caius College, University of Cambridge, Trinity St., Cambridge CB2 1TA, UK}

\date{\today}

\begin{abstract}
Predicting the densest random disc packing fraction is an unsolved paradigm problem relevant to a number of disciplines and technologies. 
One difficulty is that it is ill-defined without setting a criterion for the disorder. Another is that the density depends on the packing protocol and the multitude of possible protocol parameters has so far hindered a general solution.  
A new approach is proposed here. After formulating a well-posed form of the general protocol-independent problem for planar packings of discs, a systematic criterion is proposed to avoid crystalline hexagonal order as well as further topological order. The highest possible random packing fraction is then derived exactly: $\phi_{RCP}=0.852525...$. 
The solution is based on the cell order distribution that is shown to: (i) yield directly the packing fraction; (ii) parameterise all possible packing protocols; (iii) make it possible to define and limit all topological disorder. 
The method is further useful for predicting the highest packing fraction in specific protocols, which is illustrated for a family of simply-sheared packings that generate maximum-entropy cell order distributions. 

\end{abstract}

\maketitle

%\section{\label{sec:Intro}I. Introduction}

Understanding how particles pack is a centuries-old problem relevant to a range of applications~\cite{Toetal10}. Of particular importance is the ability to predict the highest packing fraction (PF) of disordered assemblies. Since the PF depends on the size and shape distributions of the particles, packing of identical frictionless spheres in three dimensions and discs in two have become paradigm problems, whose solutions could pave the way to solving more general cases. 

A proper solution is hindered mainly by the absence of a consensus on, and a criterion of, what is acceptable as `random'~\cite{Toetal00}. 
In particular, same-size spheres and discs tend to form crystalline regions, which increase the mean PF, and a criterion for the largest allowed such regions is required. Moreover, as discussed below, other forms of topological order also need to be avoided, an issue largely neglected in the literature. A recent proposal to address the disorder issue by defining randomness in terms of an order parameter, whose minimum is at a conceptually most random state~\cite{Toetal00}, is an attempt in the right direction, but is problematic because the choice of the order parameter is not unique. 

Another potential difficulty is that different packing protocols give rise to different PFs. For example, piling particles slowly by deposition yields PFs that depend on the deposition flux~\cite{Bletal05} and the PFs of shaken granular systems depend on the shaking frequency and amplitude~\cite{Noetal98}. Protocols cover a wide range of parameters, many of which can be varied continuously, and therefore exist in an infinite-dimensional parameter space. It then seems impossible to construct a solution that could hold for all possible protocols.

Focusing on same-size frictionless discs in the plane, a well-posed version of the problem is: 
`what is the highest possible PF out of all possible protocols?'. 
Several analytical models have been proposed, predicting PF values that range from $0.81$ to $0.89$~\cite{Stetal64,Su77,Ka80,Su80,Sh80,Sh82,Be83,Meetal10}. 
Numerical and experimental measurements unavoidably resort to specific protocols and report, on average, lower values than the analytical models~\cite{Kaetal71,ViBo72,QuTa74,BiTr84,Hietal90,Oheetal01}. Whether analytical, numerical or experimental, underlying all works is an explicit or implicit criterion for the level of disorder.

Here, I show that both the infinite-parameter space and the disorder criterion difficulties are removable for the disc problem by using the cell order distribution (COD), defined as follows. The lines joining the centres of discs in contact are the edges of a graph whose nodes are the disc centres. The graph's smallest (aka irreducible) loops are the cells. A cell's order, $k$, is the number of discs (or nodes) surrounding it. Henceforth, such a cell is called $k$-cell. The COD is the fraction of $k$-cells, $Q_k>0$ for $k = 3, 4, ...,C$ ($Q_{k>C}\equiv0$) and it has been shown to be spatially uncorrelated~\cite{Suetal20}.
The COD is key to the solution presented here because, as shown below: (i) it determines uniquely the PF, (ii) it can be used to parameterise {\it all possible protocols}, (iii) it can be used to specify the maximum allowed amount of order, (iv) when packing protocols produce large cells, the highest PF is achieved when the COD's entropy is maximal. 

The solution of the packing problem is constructed as follows. First, the COD is used to determine the PF. The use of the COD to parameterise all possible protocols is then discussed. Next, the definition of order is discussed and I argue that order goes beyond the conventional hexagonal lattice and must be extended to topological order. This discussion forms the basis for a disorder criterion. Using this criterion, the highest possible random close PF, $\phi_{RCP}$ is then derived.  
Going beyond this general solution, it is shown that this approach can be used to determine the highest PF for any specific protocol and that this PF coincides with the maximum-entropy COD.

\begin{figure}[tp]
\begin{center}
\includegraphics[width=.95\linewidth]{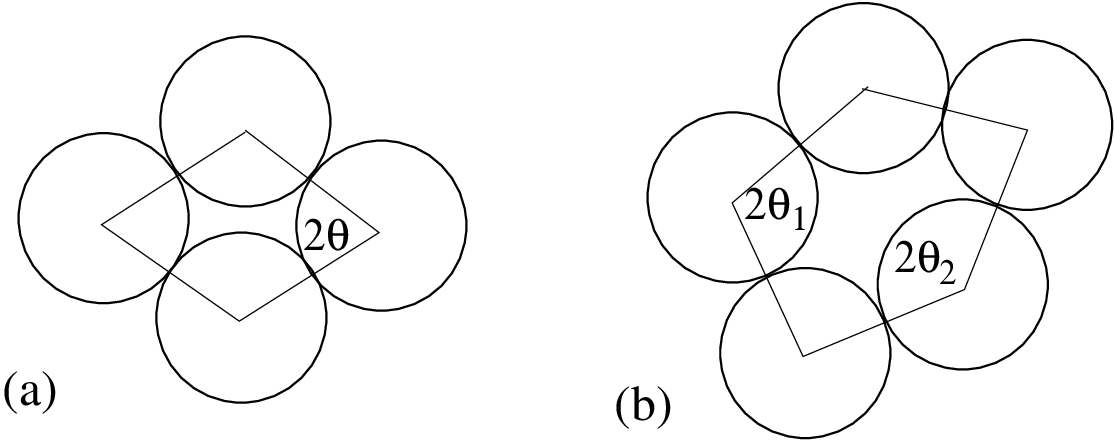}
\end{center}
\caption{ The shape of a $k$-cell depends on $k-3$ internal angles, illustrated for a $4$-cell in (a) and for a $5$-cell in (b).}
\label{S45}
\end{figure}

I consider packings of $N$ unit-diameter frictionless rigid discs, confined under vanishingly small compressive boundary stresses and presumed mechanically stable. $N$ is sufficiently large to neglect boundary effects and all the discs transmit vanishingly weak forces, which excludes rattlers~\cite{LuSt90}. A further reason for the condition on the boundary stresses will become clear later. 
The PF is the ratio of the area occupied by all the discs, $N\pi/4$, to the total area, which is the sum of the areas of all the polygonal cells,
\beq
S_{total} = N_c\sum_{k=3}^C Q_k \bar{S}_k \ ,
\label{CellAreas}
\eeq
with $N_c$ the total number of cells and $\bar{S}_k$ the mean area of all the possible $k$-cell configurations. The PF is then
\beq
\phi = \frac{\pi}{4\sum_{k=3}^C Q_k \bar{S}_k}\frac{N}{N_c} = \frac{\pi\left(\bar{k}-2\right)}{8\sum_{k=3}^C Q_k \bar{S}_k} \ .
\label{PF1}
\eeq
In (\ref{PF1}), $\bar{k}$ is the mean cell order and the relation $N/N_c = \left(\bar{k}-2\right)/2$ has been used, which is derived from Euler's relation, as shown in~\cite{SupMat21}. \\
$\bar{S}_3= \sqrt{3}/4$ is straightforward to calculate.
Cell shapes for $k>3$ are determined by $k-3$ internal angles, as exemplified in Fig. \ref{S45} for $4$- and $5$-cells -- the former depend on one angle,  $\pi/6\leq\theta\leq\pi/3$, and the latter on two, $\theta_1$ and $\theta_2$. As discussed later, the highest PF is achieved for maximum cell entropy, which implies a uniform distribution of the internal angles. For $4$-cells, this means that $P(\theta)=6/\pi$ and, using $S_4(\theta)=\sin{2\theta}$, yields
\beq
\bar{S}_4 = \frac{6}{\pi}\int\limits_{\pi/6}^{\pi/3}\sin{2\theta}d\theta = \frac{3}{\pi} \ .
\label{S3S4}
\eeq
The area of any $5$-cell is
\begin{eqnarray}
&S_5&(\theta_1,\theta_2) = \frac{\sin{2\theta_1}+\sin{2\theta_2}}{2} + \nonumber \\
&{\ }&\frac{\sqrt{8\left(\sin^2{\theta_1}+\sin^2{\theta_2}\right)-16\left(\sin^2{\theta_1}-\sin^2{\theta_2}\right)^2-1}}{4} \ .
\label{S5}
\end{eqnarray}
Averaging over $\theta_1$ and $\theta_2$ is not straightforward because their ranges are inter-dependent: $\pi/6\leq\theta_1\leq\pi/2$ and 
\beq
\frac{\pi}{2} \geq \theta_2 \geq \frac{1}{2}\arccos{\left(\frac{3}{4}+\frac{1}{2}\cos{2\theta_1}\right)} \equiv\theta_{2,min}\ .
\label{theta2}
\eeq
Taking the angle distribution again to be uniform for maximising the entropy, $\bar{S}_5$ can be calculated numerically:
\beq
\bar{S}_5 = \frac{\int\limits_{\pi/6}^{\pi/2} \left[\int\limits_{\theta_{2,min}}^{\pi/2} S_5 d\theta_2\right] d\theta_1}{\int\limits_{\pi/6}^{\pi/2} \int\limits_{\theta_{2,min}}^{\pi/2} d\theta_2 d\theta_1}=1.5481544... \ .
\label{S5Int}
\eeq
Calculating $\bar{S}_{k}$ gets increasingly cumbersome for $k>5$. However, it is clear that the higher the fractions of low-order cells the denser the packing and, therefore, calculations of $\bar{S}_{k>5}$ are not essential to solve for $\phi_{RCP}$. Nevertheless, these are required for calculating the highest PF for specific protocols that generate CODs containing high-order cells. 

The use of the COD also alleviates the infinite protocol parameter space problem -- any protocol can be  classified by the COD it produces. This reduces the protocols parameter space to only one distribution, $Q_k(C)$. The general packing problem then translates to finding the COD that gives rise to the highest random PF and the PF's value. Thus posed, it does not matter that several protocols may give rise to the same COD. 

Next, we need a disorder criterion. It is common to regard order in such packings as the occurrence of clusters of hexagonal lattices, i.e., of $3$-cells. However, largely ignored in the literature is that packings may also contain topological order, whose disruption is the highest contributor to packings entropy~\cite{AmBl17}. A pertinent example is the deformed square lattice, which is geometrically disordered but topologically ordered. Ignoring this type of order leads to misleadingly high PFs, as demonstrated below. 
The COD is further useful for limiting topological order to any desirable level because, in some systems, the conditional distribution of $Q_k$ around a cell of any order, $m$, is independent of $m$~\cite{WaMScThesis,Suetal20}. Therefore, in those systems, the probability to find a $k$-cell, with two or more other $k$-cels neighbouring it, is 
\beq
R_{kk} = Q_k\left[1 - \left(1-Q_k\right)^k - kQ_k\left(1-Q_k\right)^{k-1} \right] \ .
\label{OnekNeighbour}
\eeq
The criterion can now be set to $R_{kk}<1/k$, i.e. typically a $k$-cell has fewer than one $k$-cell neighbour, This leads to $k$-cell clusters occurrence probability that decays exponentially with size.  For example, when $R_{33}<1/3$, the probability of $3$-cell crystalline clusters of size $L$ is $<Q_3^L\left(1-Q_3\right)^{L+2}\sim e^{-\lvert\ln{\left[Q_k\left(1-Q_k\right)\right]}\rvert L}$ for $L<6$. The probability of larger clusters continues decaying with $L$, albeit at slower exponential rates. This criterion is also conveniently independent of $C$. A more general criterion could be $R_{kk}\leq\alpha/k$, with $\alpha$ chosen at will. However, the choice $\alpha=1$ is optimal, as discussed in the concluding section. 

All the ingredients are now in place to solve the random packing problem. The following numerical values are exact and can be derived to arbitrary accuracy, but are shown to six decimal accuracy, for brevity.
Using the criterion $R_{33}<1/3$ yields the highest allowed value of $Q_3$, $Q_3^{max}=0.562236$. 
Consider, first, the densest possible packing of only $3$- and $4$-cells, in which $Q_4=1-Q_3^{max}=0.437764$, yielding $\bar{k}=3.437764$. Using (\ref{PF1}) with $C=4$, gives $\phi=0.853542$. This is the highest possible PF for packings with no hexagonal order. However, this $Q_4$ yields $R_{44}=0.257783>1/4$, which means that there is a high probability of large clusters of $4$-cells. These are deformed square lattice regions and are, therefore, topologically ordered. Thus, packings of only $3$- and $4$-cells contain only regions that are ordered either one way or another and are not truly disordered -- disordered packings must include $5$-cells. This means that $\phi_{RCP}<0.853542$. 

Using (\ref{OnekNeighbour}), the highest fraction of $4$-cells satisfying $R_{44}<1/4$ is $Q_4^{max}=0.431815$. The densest packing is
$Q_3=Q_3^{max}, Q_4=Q_4^{max}, Q_5=1-Q_3^{max}-Q_4^{max}=0.005948$. Using (\ref{OnekNeighbour}) again, $R_{55}<2.08\times10^{-6}$, which is conveniently small. In this packing, $\bar{k}=3.443712$ and, using (\ref{PF1}), $\phi=0.852525$. 
This value is the highest PF possible in a truly disordered 2D packing, given the above disorder criterion, and is the solution to the random packing problem, $\phi_{RCP}=0.852525$. 

Whether or not there exist protocols that generate this ideal COD is an open question. Since experiments and numerical simulations must resort to particular protocols, this method can be used to derive the highest PF for those protocols. 
For example, applying simple quasistatic shear to the discs assembly, with specific boundary stress and shear rate, produces a specific COD, but varying those and the interparticle friction gives rise to a family of CODs. Suppose a subset of these CODs are disordered, i.e.,, the disorder criteria for $R_{33}$, $R_{44}$, and $R_{66}$ (which are the only $k$-cells that can order topologically) are satisfied. Then the highest random close PF, which this protocol can achieve, can be found by using eqs. (\ref{CellAreas}) and (\ref{PF1}) to identify the densest COD and hence the densest member of this family.  

Before continuing, the effect of rattlers has to be discussed. Rattlers are discs that occupy area, but do not participate in the force-carrying skeleton, and the definition of the PF depends on whether or not this area is taken into account. In the following, rattlers are excluded from the calculations, but including them, which increases the PF, is straightforward. 

To illustrate the procedure, consider a quasistatic cyclic shearing of the discs in the plane. The COD generated by this process can be modulated by the interparticle friction and the confining pressure and it has been shown to yield maximum-entropy CODs~\cite{Suetal20} and cell configurations, subject to the constraint of mechanical stability. The stability constraint eliminates unstable long and tortuous cells~\cite{MaBl17,Suetal20}. In any particular subset of $k$-cells, the higher the fraction of long cells the lower their mean area-to-perimeter ratio and the lower the value of $\bar{S}_k$. From eq. (\ref{PF1}), this means that such cells {\it increase} the value of $\phi$. The conclusion is that the highest PF in any physical protocol corresponds to the maximum-entropy COD, when the effect of mechanical stability is minimal. The effect of mechanical stability can be minimised by reducing the compressive boundary loads to minimum and increasing inter-particle friction~\cite{Suetal20}. This is the reason that the packing problem has been posed initially with vanishingly small compressive boundary loads.  

The maximum-entropy CODs, generated by the cyclic shear experiments in \cite{Suetal20} are exponential, as shown in~\cite{SupMat21},
\beq
Q_k = Ae^{-\lambda k} \ ,
\label{Qk}
\eeq
with $A=e^{3\lambda}\left(1-e^{-\lambda}\right)/\left[1-e^{-\lambda(C-2)}\right]$ and $\lambda$ only a function of $\bar{k}$. 
The densest disordered packing corresponds to $C=5$. Combining (\ref{Qk}) with the requirement $Q_3 = Q_3^{max} = 0.562235$, fixes the COD, as shown in~\cite{SupMat21}: $Q_4=0.289105$ and $Q_5=0.148660$. These yield $\bar{k}=3.586424$ and, using (\ref{PF1}) with $C=5$, yields 
\beq
\phi_{cs} = \frac{\pi(\bar{k} - 2)}{8\left(Q_3\bar{S}_3 + Q_4\bar{S}_4 + Q_5\bar{S}_5\right)} = 0.831007 \ .
\label{phi345}
\eeq
This is the densest possible packing that such shearing protocols can produce. Unsurprisingly, $\phi_{cs}<\phi_{RCP}$. Protocols that generate higher cell orders, $C>5$ would, unavoidably, yield lower PFs. 

To conclude, the highest possible PF has been derived analytically for a planar packing of discs under the condition that the packing is disordered geometrically and topologically. Central to the method is the cell order distribution, which makes possible: (i) direct calculation of the PF, (ii) parameterisation of all possible packing protocols, and (iii) a quantitative criterion for the disorder. The criterion, $R_{kk}\leq1/k$ for $k=3$ and $4$, has been chosen to ensure that the occurrence probability of regions of hexagonal and deformed square lattice decay exponentially with size. 
By determining the COD that corresponds to the densest possible disordered packing, the global random close packing has been found,  
$\phi_{RCP}=0.852525...$. Limiting only hexagonal order and allowing in topological order, in the form of deformed square lattices, can increase the $\phi_{RCP}$ to $0.853542$, but the contention here is that such packings are not truly disordered.
This means that calculations, simulations and physical experiments yielding higher PFs must include considerable ordered regions. In particular, predictions of $\phi_{RCP}=0.89$~\cite{Sh80,Meetal10} mean that $Q_3\geq0.898091$, $R_{33}>0.872$, and large hexagonal clusters must occur.
Whether or not there exists a physical or numerical protocol that can produce the COD that gives rise to $\phi_{RCP}$ remains an open question, Not the least because one of its objectives would be to avoid ordering.

\begin{figure}[htbp]
   \centering
   \includegraphics[width=.48\textwidth]{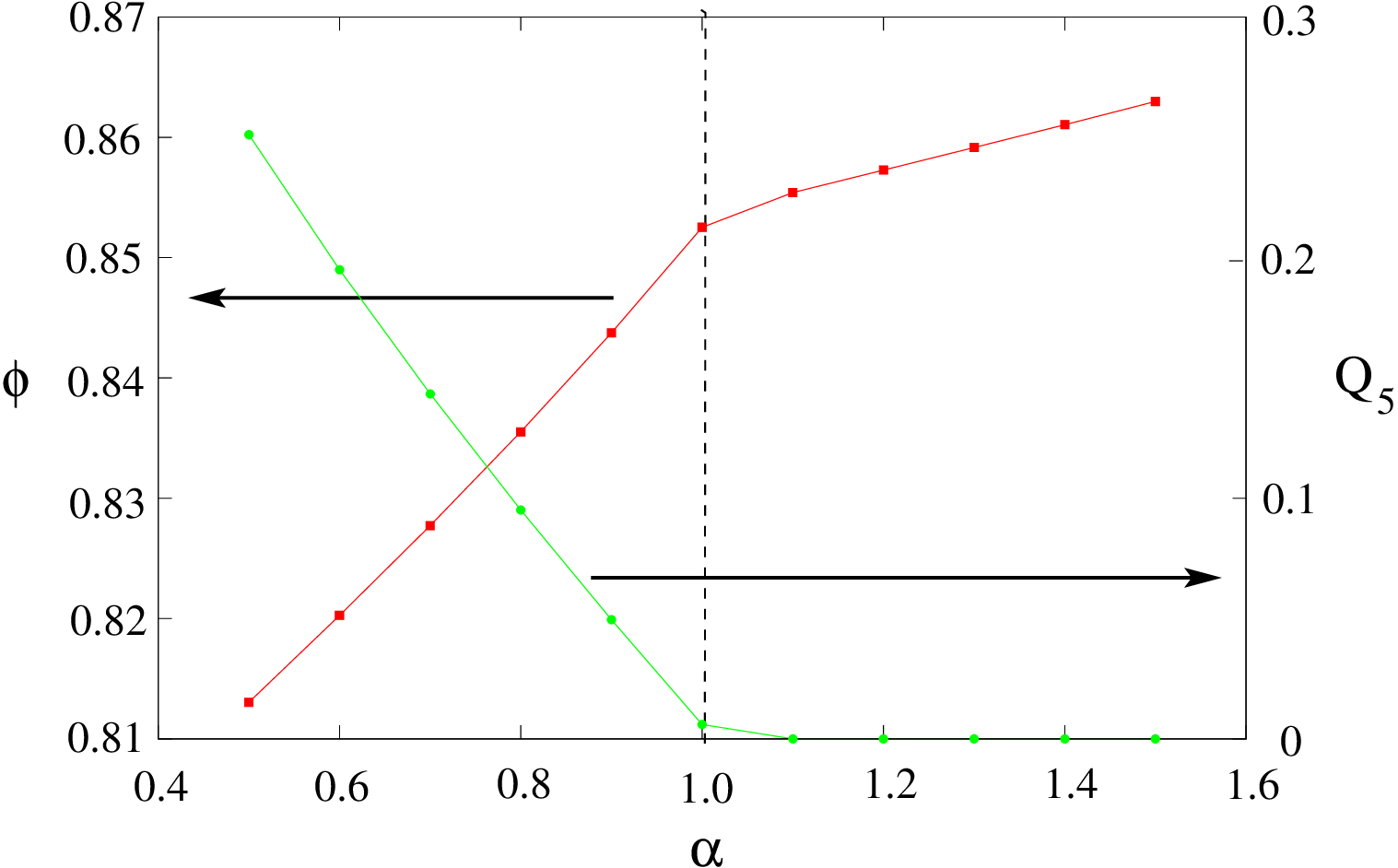}
   \caption{Increasing the disorder parameter $\alpha$, reduces $Q_5$ (right axis) and increases $\phi_{RCP}$ (left axis). At $\alpha=1$, $\phi_{RCP}=0.852525$. $Q_5$ tends to zero very slightly above $\alpha=1.0$ and well below $\alpha=1.1$ (dashed line). Beyond this point, packings contain only $3$- and $4$-cells and are therefore not truly disordered.}
   \label{RCPvsAlpha}
\end{figure}

The importance of a clear criterion for the disorder cannot be overemphasised.
The criterion chosen here can be generalised to $R_{kk}\leq\alpha/k$ for $k=3$ and $4$. As can be seen in Fig. \ref{RCPvsAlpha}, increasing $\alpha$ increases the highest PF, but this lets in more order. Three arguments motivate the choice $\alpha=1$.
(i) $\alpha>1$ means that, on average, a cell typically has more than one same-order neighbour. As a result, the occurrence probability of ordered clusters decay slowly with size already for relatively small sizes. 
(ii) Increasing $\alpha$ reduces the fraction of $5$-cells, $Q_5$, and it vanishes beyond $1.0<\alpha_c<1.1$ (see Fig. \ref{RCPvsAlpha}). As discussed, such packings are, unavoidably, topologically ordered, namely, $\alpha$ must be smaller than $\alpha_c$.
(iii) While reducing $\alpha$ below $1$ reduces further the occurrence probability of ordered clusters, this improvement is negligible and it comes at the cost of reducing the highest PF. 

The method can be used for any specific protocol producing a family of CODs. This was illustrated for experiments of cyclic shear at low confining stress that give rise to maximum-entropy CODs~\cite{Suetal20}. The highest PF that such an experiment can produce has been found to be $\phi=0.831006$. 

Other uses of the method are possible. One is for determining the highest PFs of packings whose mean coordination number is known or constrained. This information must be accompanied by knowledge of the highest possible cell order $C$. The procedure is the following: (i) use eq. (4) in the supplemental material to calculate the mean order, $\bar{k}$; (ii) derive the corresponding maximum-entropy COD corresponding to it, using the procedure leading to eq. (\ref{Qk}); (iii) determine the mean area of each cell order, $\bar{S}_k$, up to $C$ (this is the most computationally intensive step); (iv) using eq. (\ref{PF1}) calculate the highest PF.\\
Another application is to finding the highest PFs of bidisperse disc systems, which are commonly used to avoid order. This would involve the statistics of the $k$-cell configurations and their areas and should probably resort to numerical computations. \\

\begin{acknowledgments}
The author acknowledges the hospitality of the Cavendish Laboratory. Thanks go to Dr F. Rietz for spotting a sign typo.
\end{acknowledgments}


\begin{thebibliography}{99} 

\bibitem{Toetal10} S. Torquato, F.H. Stillinger, Rev. Mod. Phys. {\bf 82}, 2633 (2010) and references therein
\bibitem{Toetal00} S. Torquato, T.M. Truskett, P.G.  Debenedetti, Phys. Rev. Lett. {\bf 84}, 2064 (2000)
%\bibitem{Za08} F. Zamponi, Packings close and loose, Nature {\bf 453}, 606 (2008) (despite claims to the contrary)
\bibitem{Bletal05} R. Blumenfeld, S. F. Edwards, R. C. Ball, J. Phys.: Cond. Mat. {\bf 17}, S2481-S2487 (2005)
\bibitem{Noetal98} E. R. Nowak, J. B. Knight, E. Ben-Naim, H. M. Jaeger, S. R. Nagel, Phys. Rev. E {\bf 57}, 1971 (1998)
\bibitem{Stetal64} F.H. Stillinger, Jr., E.A. DiMarzio, R.L. Kornegay, J. Chem. Phys. {\bf 40}, 1564 (1964)
\bibitem{Su77} D. N. Sutherland, J. Colloid Interface Sci. {\bf 60}, 96 (1977)
\bibitem{Ka80} K. Kanatani, Lett. Appl. Eng. Sci. {\bf 18}, 989 (1980)
\bibitem{Su80} M. Sugiyarna, Prog. Theor. Phys. {\bf 63}, 1848 (1980)
\bibitem{Sh80} M. Shahinpoor, Powder Technol. {\bf 25}, 163 (1980)
\bibitem{Sh82} M. Shahinpoor, J. Colloid Interface Sci. {\bf 85}, 227 (1982)
\bibitem{Be83} J.G. Berryman, Phys. Rev. A {\bf 27}, 1053 (1983)
\bibitem{Meetal10} S. Meyer, C. Song, Y. Jin, K. Wang, H.A. Makse, Physica A {\bf 389}, 5137 (2010)
\bibitem{Kaetal71} H.H. Kausch, D.G. Fesko, N.W. Tschoegl, J. Colloid Interface Sci. {\bf 37}, 603 (1971)
\bibitem{ViBo72} W. M. Visscher and M. Bolsterli, Nature (London) {\bf 239}, 504 (1972)
\bibitem{QuTa74} T. J. Quickenden and G. K. Tan, J. Colloid Interface Sci. {\bf 48}, 382 (1974)
\bibitem{BiTr84} D. Bideau, J.P. Troadec, J. Phys. C: Solid State Phys. {\bf 17}, L731 (1984)
\bibitem{Hietal90} E.L. Hinrichsen, J. Feder, T. J{\o}ssang, Phys. Rev. A {\bf 41}, 4199 (1990)
\bibitem{Oheetal01} C.S. O'Hern, S.A. Langer, A.J. Liu, S.R. Nagel, Phys. Rev. Lett. {\bf 86}, 111 (2001)
\bibitem{Suetal20} X. Sun, W. Kob, R. Blumenfeld, H. Tong, Y. Wang, J. Zhang, Phys. Rev. Lett. {\bf 125}, 268005 (2020)
%\bibitem{Waetal21} C. C. Wanjura, A. Mayl\"ander, O. Marti, R. Blumenfeld, {\it The origin of detailed balance in steady states of quasi-static granular dynamics}, \red{arXiv??}
\bibitem{LuSt90} B.D. Lubachevsky, F.H. Stillinger, J. Stat. Phys. {\bf 60}, 561 (1990)
\bibitem{SupMat21} R. Blumenfeld, Supplementary material,  http://rafi.blumenfeld.co.uk/RandomClosePackingSup.pdf
\bibitem{AmBl17} S. Amitai and R. Blumenfeld, Phys. Rev. E {\bf 95}, 052905 (2017)
\bibitem{WaMScThesis} C. C. Wanjura, {\it The Structural Evolution of Granular Matter -- A Master Equation Approach}, Master's Thesis, Ulm University (2018)
\bibitem{MaBl17} T. Matsushima and R. Blumenfeld, Phys. Rev. E {\bf 95}, 032905 (2017)
\bibitem{Jo14} J.F. Jordan, {\it Computing entropy and ordering of granular materials: From description to prediction}, PhD Thesis, Imperial College London (2014)

\end{thebibliography}
\end{document}